\documentclass[12pt]{article}
\usepackage{epsfig}
\usepackage{amssymb}
\usepackage{overpic}
\usepackage{multirow}
\usepackage{dcolumn}

\def\beq{\begin{equation}}
\def\eeq#1{\label{#1}\end{equation}}
\def\eeqn{\end{equation}}

\def\beqa{\begin{eqnarray}}
\def\eeqa#1{\label{#1}\end{eqnarray}}
\def\eeqan{\end{eqnarray}}

\let\bar=\overbar

\def\etal{{\it et al.}}

\def\Dslash{\not{\hbox{\kern-4pt $D$}}}
\def\dslash{\not{\hbox{\kern-2pt $\del$}}}

\def\msb{{\bar{\ssstyle M \kern -1pt S}}}

\def\Title#1{\begin{center} {\Large {\bf #1} } \end{center}}

\newcommand{\ttbar}{\ensuremath{t\bar{t}}}

\newcommand{\etmissx}{\ensuremath{E \kern-0.6em\slash_{\rm x}}}
\newcommand{\etmissy}{\ensuremath{E \kern-0.6em\slash_{\rm y}}}

\newcommand{\ljets}{\ensuremath{\ell+{\rm jets}}}
\newcommand{\gtop}{\ensuremath{\Gamma_t}}
\newcommand{\gtwb}{\ensuremath{\Gamma_{t\rightarrow Wb}}}
\newcommand{\btwb}{\ensuremath{\mathcal B_{t\rightarrow Wb}}}
\newcommand{\btwq}{\ensuremath{\mathcal B_{t\rightarrow Wq}}}
\newcommand{\sigt}{\ensuremath{\sigma_{t-{\rm channel}}}}
\newcommand{\afb}{\ensuremath{A_{\rm fb}}}
\newcommand{\prob}{\ensuremath{\mathcal{P}}}
\newcommand{\psig}{\ensuremath{\mathcal{P}_{\rm sig}}}

\newcommand{\del}{\partial}

\newcommand{\GeV}{\ensuremath{\textnormal{GeV}}}
\newcommand{\TeV}{\ensuremath{\textnormal{TeV}}}

\newcommand{\fb}{\ensuremath{{\rm fb}^{-1}}}
\newcommand{\mtop}{\ensuremath{m_{\rm top}}}

\newcommand{\pt}{\ensuremath{p_{\rm T}}}

\newcommand{\mttbar}{\ensuremath{m_{t\bar t}}}
\newcommand{\absdy}{\ensuremath{|\Delta y|}}

\begin{document}

\Title{Top quark property measurements at the Tevatron}

\bigskip\bigskip


\begin{raggedright}  
{\it Oleg Brandt\index{Brandt, O.} on behalf of the CDF and D0 collaborations\\
II. Physikalisches Institut\\
Georg August Universit\"at G\"ottingen\\
D-37077 G\"ottingen, Germany}
\bigskip
\end{raggedright}

\begin{abstract}
We review recent measurements of the properties of the top quark by the CDF and D0 experiments: the mass and decay width of the top quark, the ratio $R=\btwb/\btwq$, the total production cross section of \ttbar\ pairs, the helicity of the $W$ boson, anomalous couplings at the $Wtb$ vertex, the asymmetry of $\ttbar$ production due to the colour charge, and the first study of the top quark polarisation. The measurements are performed using data samples of up to 8.7~\fb\ acquired by the CDF and D0 experiments in Run II of the Fermilab Tevatron $p\bar p$ collider at a centre-of-mass energy of $\sqrt s=1.96~\TeV$.
\vspace{2mm}\\
PACS {\tt 14.65.Ha} -- Top quarks.
\end{abstract}

The pair-production of the top quark was discovered in 1995 by the CDF and D0 experiments~\cite{bib:topdiscovery} at the Fermilab Tevatron proton-antiproton collider. The observation of the electroweak (EW) production of single top quarks was presented only 4 years ago~\cite{bib:singletopdiscovery}. The large top quark mass and the resulting Yukawa coupling of about $0.996\pm0.006$ suggest that the top quark could play a crucial role in EW symmetry breaking. Precise measurements of the properties of the top quark provide a crucial test of the consistency of the standard model (SM) and could hint at physics beyond the SM. The full listing of top quark measurements by the D0 experiment can be found in~Refs.~\cite{bib:toprescdf,bib:topresd0}.

At the Tevatron, top quarks are mostly produced in pairs via the strong interaction. 
In the framework of the SM, the top quark decays to a $W$ boson and a $b$ quark nearly 100\% of the time, resulting in a $W^+W^-b\bar b$ final state from top quark pair production.
Thus, $\ttbar$ events are classified according to the $W$ boson decay channels as ``dileptonic'', ``all--jets'', or ``\ljets''. The EW production of single top quarks is classified via the $s$ and $t$ channels, as well as associated $Wt$ production.

\bigskip
Currently, the Tevatron's best single measurement of \mtop\ is performed by CDF in \ljets\ final states using the so-called {\em template} method to analyse the full dataset of 8.7~\fb~\cite{bib:mtoplj_cdf}. The basic idea of the template method is to construct ``templates'', i.e.\ distributions in a set of variables~$x$, which are sensitive to \mtop, for different mass hypotheses, and extract \mtop\ by matching them to the distribution found in data, e.g.\ via a maximum likelihood fit. CDF minimises a $\chi^2$-like function to kinematically reconstruct the event for jet-parton assignments consistent with the $b$-tagging information. To extract \mtop\ and calibrate the JES {\em in-situ}, three-dimensional templates are defined in (i)~the fitted \mtop\ of the best jet-parton assignment,
(ii)~the fitted \mtop\ of the second-best assignment, and
(iii)~the fitted invariant mass of the dijet system from the hadronically decaying $W$ boson. CDF finds $\mtop=172.9~ \pm 0.7~({\rm stat}) \pm 0.8~({\rm syst})~\GeV$.

The Tevatron's most precise measurement of \mtop\ in dilepton final states is performed by D0 using 4.7~\fb\ of data~\cite{bib:mtopll_d0}. Leaving \mtop\ as a free parameter, dilepton final states are kinematically underconstrained by one degree of freedom, and the so-called neutrino weighting algorithm is applied for kinematic reconstruction. It postulates distributions in rapidities of the neutrino and the antineutrino, and calculates a weight, which depends on the consistency of the reconstructed $\vec\pt^{\,\nu\bar\nu}\equiv\vec\pt^{\,\nu}+\vec\pt^{\,\bar\nu}$ with the measured missing transverse momentum vector, versus \mtop. D0 uses the first and second moment of this weight distribution to define templates and extract \mtop. To reduce the systematic uncertainty, the {\em in situ} JES calibration in \ljets\ final states derived in Ref.~\cite{bib:mtoplj_d0} is applied, accounting for differences in jet multiplicity, luminosity, and detector ageing. 
Taking into account statistical and systematic correlations, a combination with the D0 matrix element result from the same dataset~\cite{bib:mtopllme_d0} yields  $\mtop=173.9~ \pm 1.9~({\rm stat}) \pm 1.6~({\rm syst})~\GeV$.

The combination of published and preliminary measurements of the top quark mass at the Tevatron yields $\mtop=173.2\pm0.9~$GeV~\cite{bib:mtop}. With a relative precision of slightly above 0.5\%, this is world's most precise determination of the top quark mass to date.

\bigskip
CDF directly measures the total decay width of the top quark, \gtop, using the same kinematic reconstruction and the same dataset as in Ref.~\cite{bib:mtoplj_cdf} by forming templates in reconstructed $\gtop$, and finds $\gtop=2.21^{+1.84}_{-1.11}~\GeV$~\cite{bib:width_cdf}, resulting in a top-quark lifetime of $\tau_t=(3.0^{+3.0}_{-1.3})\times10^{-25}$~s, in agreement with the SM expectation.

D0 extracts \gtop~\cite{bib:width} from the partial decay width $\gtwb$, measured using the $t$-channel cross section for single top quark production~\cite{bib:singletop}, and from the branching fraction $\btwb$, measured in $\ttbar$ events~\cite{bib:ttbar}, from up to 5.4~\fb\ of data. This extraction is made under the assumption that the EW coupling in top quark production is identical to that in the decay, and using the next-to-leading order~(NLO) calculation of the ratio $\gtwb^{\rm SM}/\sigt^{\rm SM}$, i.e.~$\gtop=\frac{\sigt}{\btwb}\times\frac{\gtwb^{\rm SM}}{\sigt^{\rm SM}}$.
Properly taking into account all systematic uncertainties and their correlations among the measurements of \gtwb\ and \sigt, D0 finds $\gtop=2.00^{+0.47}_{-0.43}~\GeV$, which translates into $\tau_t=(3.3^{+0.9}_{-0.6})\times10^{-25}$~s, in agreement with the SM expectation. This constitutes the world's most precise indirect determination of \gtop\ to date.
Furthermore, D0 uses the $t$-channel discriminant of the above measurement to extract a limit 
of $|V_{tb}|>0.81$ at 95\%~C.L., without the commonly made assumptions that $t\to Wb$ exclusively, or on the relative $s$ and $t$ channel rates.

\bigskip
The SM predicts a ratio $R\equiv\frac{\btwb}{\btwq}=\frac{|V_{tb}|^2}{|V_{td}|^2+|V_{ts}|^2+|V_{tb}|^2}$ of almost unity. CDF measures the total production cross section of \ttbar\ pairs and simultaneously extracts $R$ in \ljets\ final states using a dataset of 8.7~\fb. The extraction of $R$ is performed with a template fit in bins of one or two $b$-tags using events with 3, 4, and 5 or more jets, for which a different relative population is predicted for varying $R$. Taking into account systematic uncertainties as nuisance parameters, CDF finds $\sigma_{\ttbar}=7.5\pm0.95~$pb and $R=0.94\pm0.10$~\cite{bib:xsec_cdf} by performing a binned maximum likelihood fit. The measurement can be translated into a limit on $|V_{tb}|>0.89$ at 95\% CL.

\bigskip
In the SM, the top quark decays into a $W$ boson and a $b$ quark with a probability of $>99.8\%$, where the on-shell $W$ boson can be in a left-handed, longitudinal, or right-handed helicity state. A NLO calculation in the SM predicts $f_-=0.301,\,f_0=0.698,$ and $f_+=4.1\times10^{-4}$, respectively. A deviation from the SM expectation could indicate a contribution from new physics. CDF measures the $f_0$ and $f_+$ helicity fractions in \ljets\ final states using 8.7~\fb\ of data using the matrix element technique. This approach calculates the probability that a given event, characterised by a set of measured observables $x$, comes from the \ttbar\ production given $f_0$ and $f_+$, or from a background process: 
$\prob_{\rm evt}(x)\propto f\prob_{\rm sig}(x,f_0,f_+)+(1-f)\prob_{\rm bgr}(x)$.
The dependence on $\mtop$ is explicitly introduced by calculating $\psig$ using the differential cross section ${\rm d}\sigma(f_0,f_+)\propto|\mathcal{M}_{\ttbar}|^2(f_0,f_+)$, where $\mathcal{M}_{\ttbar}$ is the leading order (LO) matrix element for \ttbar\ production.
CDF measures $f_0=0.73\pm0.09$ and $f_+=-0.05\pm0.07$~\cite{bib:whel_cdf}, in agreement with the SM expectation.


\bigskip
The SM provides a purely left-handed vector coupling at the $Wtb$ vertex, while the most general and lowest-dimension effective Lagrangian 
allows a right-handed vector coupling $f^R_V$ as well as tensor couplings $f^L_T$ and $f^R_T$. D0 extracts limits on those anomalous couplings from single top production in 5.4~\fb\ using both shapes of kinematic distributions as well as the overall and $s$ versus $t$~channel event rates~\cite{bib:anomal_st}. This was done under the assumption of real, i.e.\ $CP$-conserving couplings and a spin~1/2\linebreak top quark predominantly decaying to $Wb$. 
The results are shown in Table~\ref{tab:anomal}.\\
Furthermore, D0 exploits that anomalous couplings at the $Wtb$ vertex will alter the rates of the three helicity states of $W$ bosons in $\ttbar$ decays, and combines the above analysis with the one in Ref.~\cite{bib:whel_d0}
to obtain improved limits~\cite{bib:anomal}, shown in Table~\ref{tab:anomal}.

\begin{table}[htbp]
\centering
\begin{tabular}{l|ccc}
\hline
Scenario & $W$ helicity only & single top only & combination \\
\hline 
$|f_V^R|^2$ & 0.62     & 0.89     & 0.30 \\
$|f_T^L|^2$ & 0.14     & 0.07     & 0.05 \\
$|f_T^R|^2$ & 0.18     & 0.18     & 0.12 \\
\hline
\end{tabular}
\caption{
  \label{tab:anomal}
  Observed upper limits on anomalous $Wtb$ couplings at 95\% C.L. from $W$~boson helicity assuming $f_V^L=1$, from the analysis of single top events, and their combination, for which no assumption on $f_V^L$ is made, as measured by D0 using 5.4~\fb.
}
\end{table}


\bigskip
In the SM, the pair production of top quarks in $p\bar p$ collisions, a $CP$ eigenstate, is symmetric at LO under charge conjugation. NLO calculations 
predict a small forward-backward asymmetry $\afb$ of the order of 5\% in the $\ttbar$ rest frame. It is due to a negative contribution from the interference of diagrams for initial and final state radiation, and a (larger) positive contribution from the interference of box and tree-level diagrams. This experimental situation is unique to the Tevatron. 
A convenient observable for the Tevatron is $\afb\equiv\frac{N^{\Delta y>0}-N^{\Delta y<0}}{N^{\Delta y>0}+N^{\Delta y<0}},$ where $\Delta y\equiv y_t-y_{\bar t}$, $y_t$~($y_{\bar t}$) is the rapidity of the $t$~($\bar t$) quark. 
Another common observable does not depend on a full reconstruction of the \ttbar\ system: $\afb^\ell\equiv\frac{N^{q_\ell y_\ell>0}-N^{q_\ell y_\ell<0}}{N^{q_\ell y_\ell>0}+N^{q_\ell y_\ell<0}}$, where $q_\ell$ is the lepton charge.

\begin{figure}[htb]
\centering
\begin{overpic}[width=0.48\textwidth]{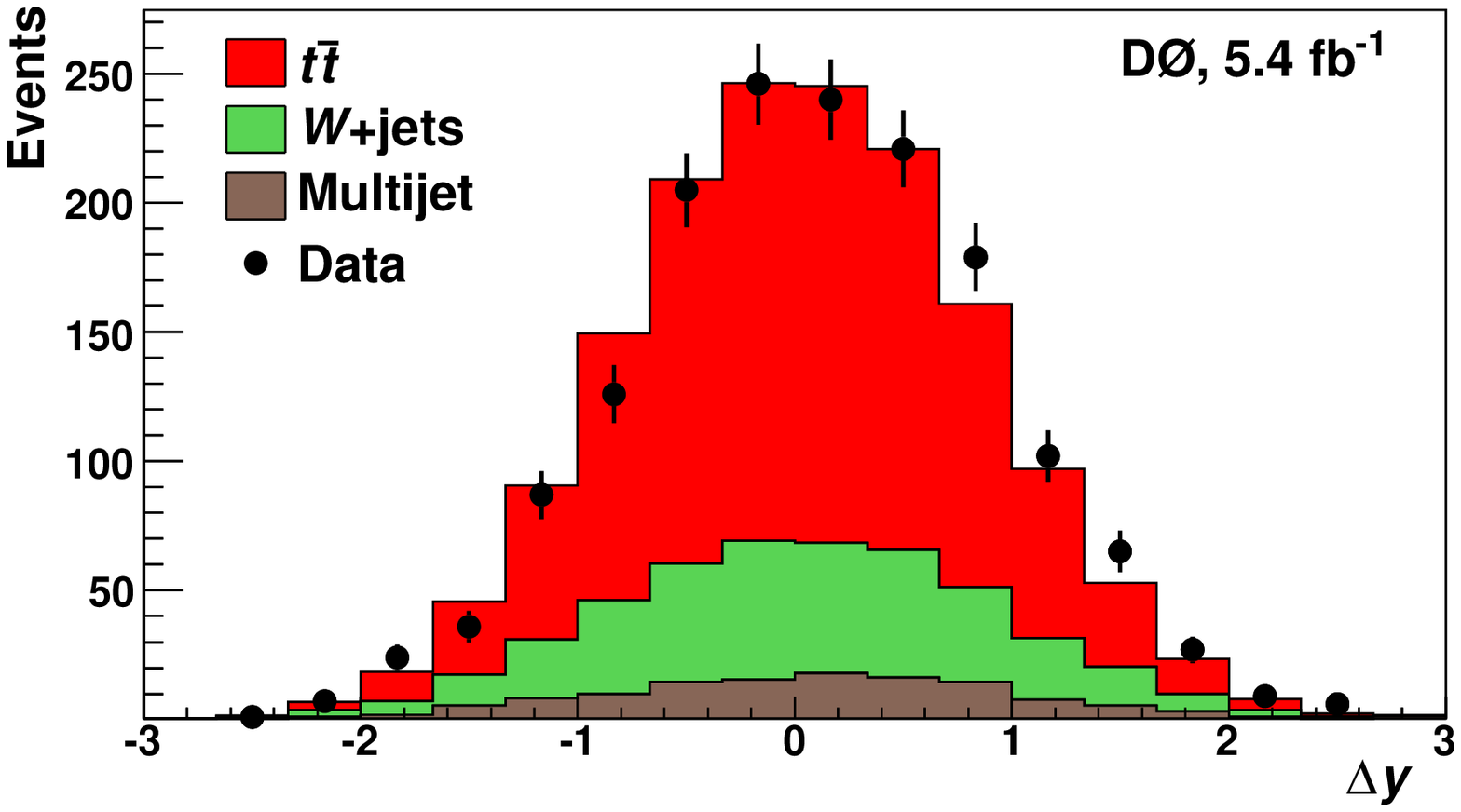}
\put(-1,2){(a)}
\end{overpic}
\quad
\begin{overpic}[width=0.48\textwidth]{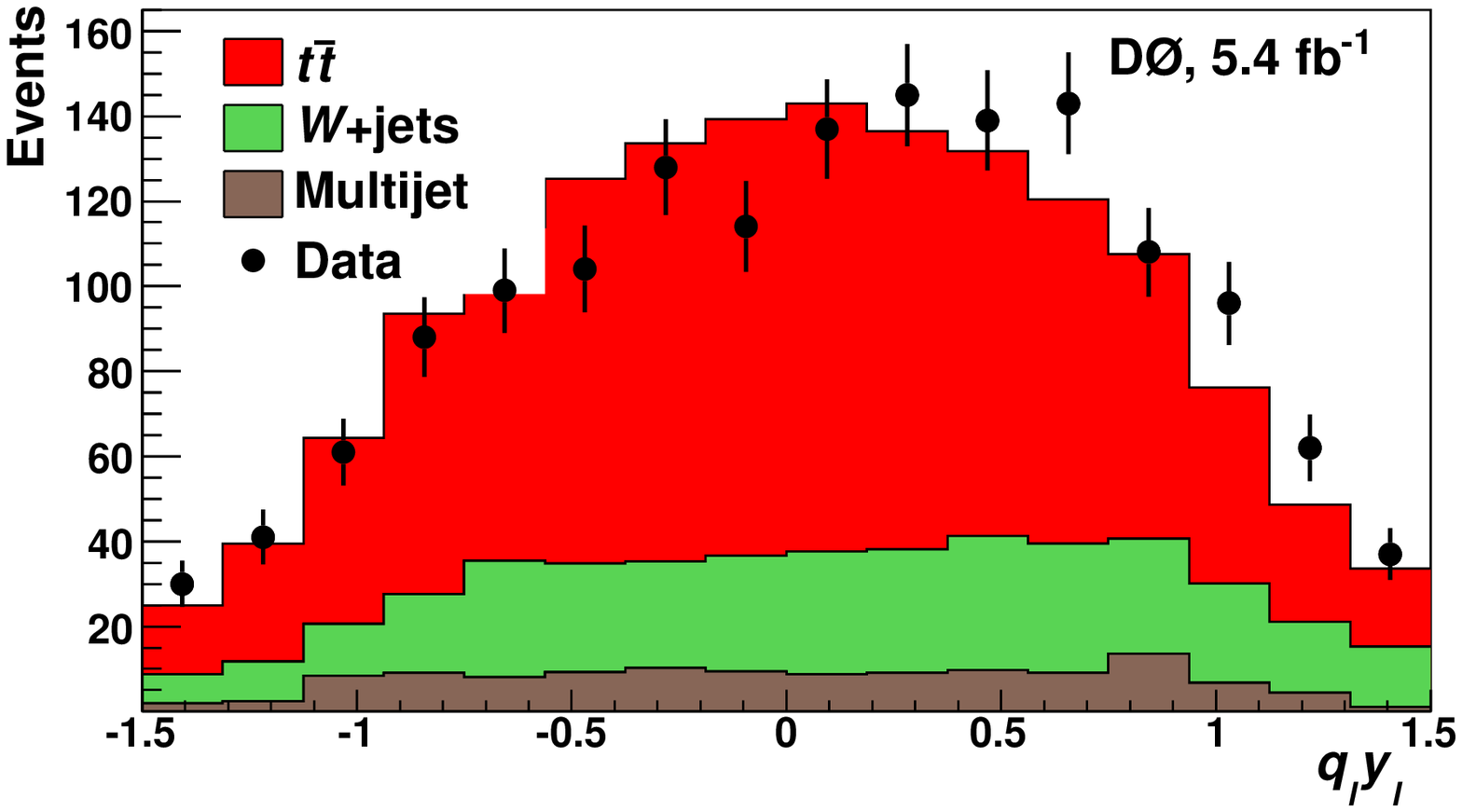}
\put(-1,2){(b)}
\end{overpic}
\caption{
\label{fig:afb}
{\bf(a)} The reconstructed distribution in $\Delta y$ in \ljets\ final states using 5.4~\fb. 
{\bf(b)} The reconstructed distribution in charge-signed lepton rapidity $q_\ell y_\ell$.
}
\end{figure}


%
D0 measures \afb\ in the $\ttbar$ rest frame in \ljets\ final states on a dataset corresponding to 5.4~\fb\ using $\ttbar$ event candidates fully reconstructed with a kinematic fitter, and finds $\afb=9.2\%\pm3.7\%$ at the reconstruction level~\cite{bib:afblj_d0}. The result, shown in Fig.~\ref{fig:afb}~(a), is about 1.9 standard deviations (SD) away from the {\sc mc@nlo}~\cite{bib:mcnlo} prediction of $2.4\pm0.7\%$. After correcting for detector acceptance and resolution we find $\afb=19.6\pm6.5\%$, 2.4~SD away from the {\sc mc@nlo} prediction of $5.0\pm0.1\%$. In addtion, D0 measures \afb\ in various subsamples
defined by $\mttbar\lessgtr450~\GeV$ and by $\absdy\lessgtr1.0$.
No statistically significant dependencies are found. Furthermore, D0 measures the lepton-based asymmetry and finds $\afb^{\ell}=14.2\pm3.8\%$ and $15.2\pm4.0\%$ at reconstruction and parton level, respectively, while {\sc mc@nlo} predicts  $\afb^{\ell}=0.8\pm0.6\%$ and $2.1\pm0.1\%$. 

D0 also measures the colour charge asymmetry in dilepton final states on a dataset corresponding to 5.4~\fb. For this measurement, robust lepton-based observables like $\afb^\ell$ are used, which do not require the full kinematic reconstruction of the events, and which can be easily corrected for background contributions and detector effects. D0 finds $\afb^\ell=3.3 \pm 6.1\%$ at reconstruction level, which translates into $\afb^\ell = 5.8 \pm 5.3\%$ after correcting for background contributions and detector effects. The result is consistent with the NLO prediction  including NLO EW corrections~\cite{bib:afb_theo} of $\afb^\ell=4.7 \pm 0.1\%$.

The D0 collaboration combines the two above results properly accounting for correlations and systematic uncertainties, and finds $\afb^\ell=11.8\pm3.2\%$. The corresponding theoretical prediction for the combination of both channels is $\afb^\ell=4.7 \pm 0.1\%$, including NLO EW corrections.

The above results display some tension with the NLO SM prediction. This may indicate a contribution from new physics, but may as well be due to contributions at higher orders in $\alpha_s$ within the SM. Similar findings by CDF are discussed in Ref.~\cite{bib:afb_cdf}.

\bigskip
I presented recent measurements of key properties of the top quark by the CDF and D0 experiments, most in good agreement with SM expectations. The forward-backward  asymmetry $\afb$ of $\ttbar$ production, as measured by D0, displays tension between the measurement and the SM NLO calculations. We look forward to updates with the full dataset of 9.7~\fb\ in the near future.

\bigskip
I would like to thank my collaborators from the CDF and D0 experiments for their help in preparing this article. I also thank the staffs at Fermilab and collaborating institutions, as well as the CDF and D0 funding agencies.

\end{document}